\documentstyle[12pt]{article}
\newcommand{\nc}{\newcommand}
\nc{\postscript}[2]
{\setlength{\epsfxsize}{#2\hsize}\centerline{\epsfbox{#1}}}
\nc{\bg}{B. Grzadkowski}
\nc{\non}{\nonumber}
\def\dps{\displaystyle}
\def\mib#1{\mbox{\boldmath $#1$}}
\def\sla#1{\mbox{$#1\!\!\scriptstyle{/}$}}

\def\bra#1{\langle #1 |} \def\ket#1{|#1\rangle}
\def\vev#1{\langle #1\rangle}

\nc{\barx}{\bar{x}}\nc{\pbarn}{\;\hbox {pb}}\nc{\fbarn}{\;\hbox {fb}}
\nc{\hc}{\hbox {h.c.}} \nc{\re}{\hbox {Re}} 
\nc{\mev}{\hbox {MeV}} \nc{\gev}{\;\hbox {GeV}}
\def\gesim{\lower0.5ex\hbox{$\:\buildrel >\over\sim\:$}}
\def\lesim{\lower0.5ex\hbox{$\:\buildrel <\over\sim\:$}}
\nc{\prd}[3]{{\it Phys.\ Rev.}\ {{\bf D{#1}} (#2), #3}}
\nc{\prl}[3]{{\it Phys.\ Rev.\ Lett.}\ {{\bf {#1}} (#2), #3}}
\nc{\plb}[3]{{\it Phys.\ Lett.}\ {{\bf B{#1}} (#2), #3}}
\nc{\npb}[3]{{\it Nucl.\ Phys.}\ {{\bf B{#1}} (#2), #3}}
\nc{\ptp}[3]{{\it Prog.\ Theor.\ Phys.}\ {{\bf {#1}} (#2), #3}}
\nc{\zfp}[3]{{\it Z.\ Phys.}\ {{\bf C{#1}} (#2), #3}}
\nc{\epj}[3]{{\it Eur.\ Phys.\ J.}\ {{\bf C{#1}} (#2), #3}}
\nc{\mpla}[3]{{\it Mod.\ Phys.\ Lett.}\ {{\bf A{#1}} (#2), #3}}
\nc{\rmp}[3]{{\it Rev.\ Mod.\ Phys.}\ {{\bf {#1}} (#2), #3}}
\nc{\ijmpa}[3]{{\it Int.\ J.\ of\ Mod.\ Phys.}\
               {{\bf A{#1}} (#2), #3}}
\nc{\ttbar}{t\bar{t}}         \nc{\bbbar}{b\bar{b}}
\nc{\tanb}{\tan \beta}        \nc{\twbdec}{t\to W^+ b}
\nc{\tbwbdec}{\bar{t}\to W^- \bar{b}}
\nc{\epem}{e^+e^-}            \nc{\eett}{\epem \to \ttbar}
\nc{\sigeett}{\sigma_{e\bar{e}\to\ttbar}}
\nc{\wpwm}{W^+W^-}            \nc{\tbar}{\bar{t}}
\nc{\bbar}{\bar{b}}           \nc{\wpp}{W^+}
\nc{\mt}{m_t}    \nc{\mts}{m_t^2}   \nc{\mw}{m_W}    \nc{\mws}{m_W^2}
\nc{\mz}{m_Z}    \nc{\mzs}{m_Z^2}
\nc{\ttbardec}{\ttbar \to W^+W^-\bbbar}
\nc{\wwbb}{W^+W^-\bbbar}      \nc{\sm}{SM}
\nc{\cw}{\cos\theta_W}        \nc{\sw}{\sin\theta_W}
\nc{\sws}{\sin^2\theta_W}     \nc{\sig}{\sigma_{tot}}
\nc{\lp}{{\ell}^+}              \nc{\lm}{{\ell}^-}
\nc{\epsl}{\epsilon_L}        \nc{\cp}{C\!P}
\nc{\splus}{s_+}       \nc{\smin}{s_-}        \nc{\eps}{\epsilon}
\nc{\psp}{Ps_+}        \nc{\psm}{Ps_-}        \nc{\lsp}{ls_+}
\nc{\lsm}{ls_-}        \nc{\sss}{s_+s_-}      \nc{\m}{m_t}
\nc{\mq}{m_t^2}        \nc{\mr}{\frac{1}{\m}} \nc{\av}{A_{\gamma}}
\nc{\bv}{B_{\gamma}}   \nc{\az}{A_Z}          \nc{\bz}{B_Z}
\nc{\avs}{A_{\gamma}^2}\nc{\azs}{A_Z^2}       \nc{\bzs}{B_Z^2}
\nc{\dav}{\delta \! A_{\gamma}}   \nc{\dbv}{\delta \! B_{\gamma}}
\nc{\dcv}{\delta C_{\gamma}}      \nc{\ddv}{\delta \! D_{\gamma}}
\nc{\daz}{\delta \! A_Z}          \nc{\dbz}{\delta \! B_Z}
\nc{\dcz}{\delta C_Z}             \nc{\ddz}{\delta \! D_Z}
\nc{\dev}{\delta \! E_{\gamma}}   \nc{\dez}{\delta \! E_Z}
\nc{\dfv}{\delta \! F_{\gamma}}   \nc{\dfz}{\delta \! F_Z}
\nc{\rdav}{{\rm Re}(\delta \! A_{\gamma}) \:}
\nc{\rdbv}{{\rm Re}(\delta \! B_{\gamma}) \:}
\nc{\rdcv}{{\rm Re}(\delta C_{\gamma}) \:}
\nc{\rddv}{{\rm Re}(\delta \! D_{\gamma}) \:}
\nc{\rdaz}{{\rm Re}(\delta \! A_Z) \:}
\nc{\rdbz}{{\rm Re}(\delta \! B_Z) \:}
\nc{\rdcz}{{\rm Re}(\delta C_Z) \:}
\nc{\rddz}{{\rm Re}(\delta \! D_Z) \:}
\nc{\idav}{{\rm Im}(\delta \! A_{\gamma}) \:}
\nc{\idbv}{{\rm Im}(\delta \! B_{\gamma}) \:}
\nc{\idcv}{{\rm Im}(\delta C_{\gamma}) \:}
\nc{\iddv}{{\rm Im}(\delta \! D_{\gamma}) \:}
\nc{\idaz}{{\rm Im}(\delta \! A_Z) \:}
\nc{\idbz}{{\rm Im}(\delta \! B_Z) \:}
\nc{\idcz}{{\rm Im}(\delta C_Z) \:}
\nc{\iddz}{{\rm Im}(\delta \! D_Z) \:}
\nc{\cz}{(1+v_e^2)d\:\!'^2}         \nc{\ci}{v_ed\:\!'}
\nc{\ccz}{v_ed\:\!'^2}              \nc{\cci}{d\:\!'}
\nc{\lspace}{\;\;\;\;\;\;\;\;\;\;}  \nc{\llspace}{\lspace \lspace}
\nc{\beq}{\begin{equation}}   \nc{\eeq}{\end{equation}}
\nc{\bea}{\begin{eqnarray}}   \nc{\eea}{\end{eqnarray}}
\nc{\baa}{\begin{array}}      \nc{\eaa}{\end{array}}
\nc{\bit}{\begin{itemize}}    \nc{\eit}{\end{itemize}}
\nc{\ben}{\begin{enumerate}}  \nc{\een}{\end{enumerate}}
\nc{\bce}{\begin{center}}     \nc{\ece}{\end{center}}
\nc{\ocal}{{\cal O}}
\begin{document}
\pagestyle{empty} \setlength{\footskip}{2.0cm}
\setlength{\oddsidemargin}{0.5cm} \setlength{\evensidemargin}{0.5cm}
\renewcommand{\thepage}{-- \arabic{page} --}
\def\mib#1{\mbox{\boldmath $#1$}}
\def\bra#1{\langle #1 |}      \def\ket#1{|#1\rangle}
\def\vev#1{\langle #1\rangle} \def\dps{\displaystyle}
\nc{\tb}{\stackrel{{\scriptscriptstyle (-)}}{t}}
\nc{\bb}{\stackrel{{\scriptscriptstyle (-)}}{b}}
\nc{\fb}{\stackrel{{\scriptscriptstyle (-)}}{f}}
\nc{\pp}{\gamma \gamma}
\nc{\pptt}{\pp \to \ttbar}
   \def\thebibliography#1{\centerline{REFERENCES}
     \list{[\arabic{enumi}]}{\settowidth\labelwidth{[#1]}\leftmargin
     \labelwidth\advance\leftmargin\labelsep\usecounter{enumi}}
     \def\newblock{\hskip .11em plus .33em minus -.07em}\sloppy
     \clubpenalty4000\widowpenalty4000\sfcode`\.=1000\relax}\let
     \endthebibliography=\endlist
   \def\sec#1{\addtocounter{section}{1}\section*{\hspace*{-0.72cm}
     \normalsize\bf\arabic{section}.$\;$#1}\vspace*{-0.3cm}}
\vspace*{-1cm}
\begin{flushright}
$\vcenter{
\hbox{IFT-19-02}
\hbox{TOKUSHIMA Report}
\hbox{(hep-ph/0208079)}
}$
\end{flushright}

\vskip 1cm
\begin{center}
{\large\bf Decoupling of Anomalous Top-Quark-Decay Vertices}

\vskip 0.15cm
{\large\bf in Angular Distribution of Secondary Particles}
\end{center}

\vspace*{1cm}
\begin{center}
\renewcommand{\thefootnote}{\alph{footnote})}
{\sc Bohdan GRZADKOWSKI$^{\:1),\:}$}\footnote{E-mail address:
\tt bohdan.grzadkowski@fuw.edu.pl}\ and\
{\sc Zenr\=o HIOKI$^{\:2),\:}$}\footnote{E-mail address:
\tt hioki@ias.tokushima-u.ac.jp}
\end{center}

\vspace*{1cm}
\centerline{\sl $1)$ Institute of Theoretical Physics,\ Warsaw
University}
\centerline{\sl Ho\.za 69, PL-00-681 Warsaw, POLAND}

\vskip 0.3cm
\centerline{\sl $2)$ Institute of Theoretical Physics,\
University of Tokushima}
\centerline{\sl Tokushima 770-8502, JAPAN}

\vspace*{2cm}
\centerline{ABSTRACT}

\vspace*{0.4cm}
\baselineskip=20pt plus 0.1pt minus 0.1pt
Angular distribution of a secondary particle from top-quark decays is
studied in a simple and general manner, paying careful attention to
how relevant the top-quark production mechanism is. The conditions
that lead to the distribution free from any possible anomalous
top-quark decay interactions are specified. Some approximations
adopted in earlier papers are relaxed and their relevance is
discussed.
\vspace*{0.4cm} \vfill



\newpage
\renewcommand{\thefootnote}{\sharp\arabic{footnote}}
\pagestyle{plain} \setcounter{footnote}{0}
\baselineskip=21.0pt plus 0.2pt minus 0.1pt
Top-quark interactions could provide relevant information on physics
beyond the Standard Model (SM) because of its huge mass. For instance
the top-quark Yukawa couplings are expected to be enhanced comparing
to those for lighter fermions and therefore precise tests of
top-quark interactions could either reveal new scalar degrees of
freedom or limit possible extensions of the SM Higgs sector. Since
future high-energy accelerators like NLC/LHC will operate as
factories of top quarks, a lot of attention has been paid to study
their production mechanisms (for a review, see \cite{Atwood:2001tu}
and the reference list there).

Only anomalous $t\bar{t}\gamma$, $t\bar{t} Z$ and $t\bar{t}g$
couplings have usually been considered in those studies, however
there is a priori no good reason to assume that the top-quark decay
is properly described by the SM couplings. Therefore in a series of
papers (see e.g.
\cite{Grzadkowski:1996kn,Grzadkowski:2000iq,Grzadkowski:2000nx}) we
have performed analysis of top-quark decay products assuming the most
general couplings both for the production and the decay.

In Ref.\cite{Grzadkowski:2000iq} we found that the angular
distribution of the final leptons in $\epem \to t\bar{t} \to
{\ell}^{\pm}\cdots$ is not sensitive to modification of the SM
$V\!-\!A$ decay vertex. The same conclusion was also reached by
Rindani \cite{Rindani:2000jg} through an independent calculation.
We usually suffer from too many parameters to be determined while
testing top-quark couplings in a general model-independent way.
Therefore, a distribution insensitive to a certain class of
non-standard form factors is obviously a big advantage as it
increases expected precision for the determination of other remaining
relevant couplings \cite{Grzadkowski:2000nx}. Furthermore in
Ref.\cite{Grzadkowski:2001tq} we have noticed that this phenomenon
appears not only in the process $\epem \to t\bar{t}\to {\ell}^{\pm}
\cdots$, but also in any $t\bar{t}$ production process.

There, however, we have limited ourselves to semileptonic SM-like
decays $t\to bW \to b{\ell}\nu$ from $t\bar{t}$ pair production and
used the explicit decay-width formula. The aim of this report is to
complete full generalization of those preceding works. That is,
we intend to relax these conditions studying general top-quark
productions and its decays as model-independently as possible.
Eventually we will specify the necessary conditions for decoupling
of the anomalous top-quark-decay effects in the angular
distributions of secondary particles from the top-quark decays.

Let us consider a general top-quark production process $1 + 2 \to
t + \cdots$ followed by its decay $t \to f + \cdots$ where $f$
denotes the secondary particle that we are going to
observe.\footnote{Note that we are not limiting ourselves here to the
   SM-like decays with $f={\ell}^+, b$ which we have considered in
   earlier papers.}\
Following our preceding works, we assume $\sqrt{s}$ to be much
higher than the threshold energy.
Since the ratio of the top-quark width ${\mit\Gamma}_t$ to its mass
$m_t$ is of the order of $10^{-2}$ we can adopt the narrow-width
approximation in the open-top region. Then one can apply the
Kawasaki-Shirafuji-Tsai formula \cite{technique} in order to determine
the $f$ distribution:\footnote{In the threshold region, this formula
   might be no longer valid due to large corrections. For example,
   non-factorizable QCD corrections appear at the level of 10 \%
   in $e\bar{e}\to t\bar{t}\to {\ell}^{\pm}X$ \cite{Peter:1997rk}
   (see also \cite{Hoang:2000yr} and references therein for studies
   in the threshold region).}
\begin{eqnarray}
&&\frac{d\sigma}{d\tilde{\mib{p}}_f} \equiv
\frac{d\sigma}{d\tilde{\mib{p}}_f}(1 + 2 \to t + \cdots \to f +
\cdots)  \non\\
&&\phantom{\frac{d\sigma}{d\tilde{\mib{p}}_f}}
=2\int d\tilde{\mib{p}}_t\frac{d\sigma}{d\tilde{\mib{p}}_t}(s_t = n)
\frac1{{\mit\Gamma}_t}\frac{d{\mit\Gamma}}{d\tilde{\mib{p}}_f}
=2B_f\int d\tilde{\mib{p}}_t
\frac{d\sigma}{d\tilde{\mib{p}}_t}(s_t = n)
\frac1{{\mit\Gamma}}\frac{d{\mit\Gamma}}{d\tilde{\mib{p}}_f}.
\label{KST1}
\end{eqnarray}
Here $d\tilde{\mib{p}}$ denotes the Lorentz-invariant phase-space
element $d\mib{p}/[\,(2\pi)^3 2p^0\,]$,
$d{\mit\Gamma}/d\tilde{\mib{p}}_f$ is the spin-averaged top-quark
width
\[
\frac{d{\mit\Gamma}}{d\tilde{\mib{p}}_f} \equiv
\frac{d{\mit\Gamma}}{d\tilde{\mib{p}}_f}(t\to f + \cdots),
\]
$B_f\equiv {\mit\Gamma}/{\mit\Gamma}_t$, and
$d\sigma(s_t = n)/d\tilde{\mib{p}}_t$ is the single-top-quark
inclusive cross section
\[
\frac{d\sigma}{d\tilde{\mib{p}}_t}(s_t = n) \equiv
\frac{d\sigma}{d\tilde{\mib{p}}_t}(1 + 2 \to t + \cdots \,;\:s_t=n)
\]
with the polarization vector $s_t$ being replaced with the so-called
``effective polarization vector" $n$
\begin{equation}
n_\mu = -\Bigl[\:g_{\mu\nu}-\frac{{p_t}_\mu{p_t}_\nu}{m_t^2}\:\Bigr]
{\sum_{\rm spin}\dps{\int}
d{\mit\Phi}\:\bar{B}{\mit\Lambda}_+\gamma_5 \gamma^\nu B \over
\sum_{\rm spin}\dps{\int}d{\mit\Phi}\:\bar{B}{\mit\Lambda}_+ B},
\label{n-vec}
\end{equation}
where the spinor $B$ is defined such that the matrix element for
$t(s_t)\to f +\cdots$ is expressed as $\bar{B}u_t(p_t,s_t)$,
${\mit\Lambda}_+\equiv \sla{p}_t +m_t$, $d{\mit\Phi}$ is the relevant
final-state phase-space element, and $\sum_{\rm spin}$ denotes the
appropriate spin summation.

The angular distribution of $f$ shall be calculated based on
eq.(\ref{KST1}). Since $d{\mit\Gamma}/d\tilde{\mib{p}}_f$ is a
Lorentz-invariant quantity depending only on $p_t$ and $p_f$, the
distribution is a function of $p_t \, p_f$ alone
\begin{equation}
d{\mit\Gamma}/d\tilde{\mib{p}}_f = F(\xi), \label{F}
\end{equation}
where $\xi\equiv p_t\, p_f = E_t E_f (1-\beta\cos\theta_{tf})$,
$\beta=|\mib{p}_t|/E_t$ and we have neglected the mass of $f$.
Integrating over $\mib{p}_f$ we find:
\beq
{\mit\Gamma}=\frac1{2(2\pi)^2\beta E_t}\int dE_f d\xi\, F(\xi)=
\frac1{(2\pi)^2 m_t^2}\int d\xi\, \xi\, F(\xi),
\label{width}
\eeq
where we used the following constraint on $E_f$ for a fixed $\xi$ in
the $E_f$ integration:
\begin{equation}
\frac{\xi}{E_t(1+\beta)} \leq E_f \leq \frac{\xi}{E_t(1-\beta)}.
\end{equation}
On the other hand, the $f$ angular distribution is
\beq
\frac{d\sigma}{d{\mit\Omega}_f}
=\frac{B_f}{(2\pi)^3}
\int dE_f E_f \int d\tilde{\mib{p}}_t
\frac{d\sigma}{d\tilde{\mib{p}}_t}(s_t=n)
\frac1{{\mit\Gamma}}F(\xi).  \label{Pre-ang}
\eeq
Here the polarization-vector $n$ can in general depend on $E_f$ and
the integration over $E_f$ cannot be performed before explicit
calculation of $d\sigma(s_t=n)/d\tilde{\mib{p}}_t $.

However, if the vector $n$ is free from $E_f$, we can perform the
$E_f$ integration independently of the production mechanism since
$d\sigma/d\tilde{\mib{p}}_t$ can depend on $E_f$ only through $n$
inserted instead of $s_t$. Let us briefly discuss possible dependence
of $n$ on $E_f$. From the definition (\ref{n-vec}), $n$ is a linear
combination of $p_t$ and $p_f\,$: $n=a \,p_t + b \,p_f$. Since $n\,
p_t=0$ we obtain
\beq
n = \alpha^f\Bigl(\frac{m_t}{p_t \, p_f}p_f - \frac{p_t}{m_t}\Bigr),
\label{pol}
\eeq
where we have introduced the depolarization factor $\alpha^f \equiv
-a\,m_t$. We also have the following inequality from eq.(\ref{n-vec})
\beq
-1 \leq n^2 \leq 0\ \ \ \Longrightarrow\ \ \
-1 \leq \alpha^f \leq 1, \label{n-2}
\eeq
which can be easily proven in the top-quark rest frame via a
generalized triangle inequality. Since we assumed that $f$ is
massless, $E_f$ drops out in $p_f/(p_t \, p_f)$ in eq.(\ref{pol}) and
the above inequality constrains the $E_f$ dependence of $\alpha^f$.
This discussion is never a proof that $n$ is always $E_f$
independent, but eq.(\ref{n-2}) is a strong constraint and it will
not be unreasonable to consider an $E_f$-independent $n$ vector.

Therefore, let us now temporarily assume that $n$ is not a function
of $E_f$. Then we can  carry out the $E_f$ integration in
eq.(\ref{Pre-ang}):
\begin{equation}
\frac{d\sigma}{d{\mit\Omega}_f}
=\frac{B_f}{(2\pi)^3}
\int d\tilde{\mib{p}}_t \frac{d\sigma}{d\tilde{\mib{p}}_t}(s_t=n)
\frac1{{\mit\Gamma}}\int dE_f E_f F(\xi).
\end{equation}
Since eq.(\ref{width}) gives
\[
\int dE_f E_f F(\xi)=\frac1{E_t^2(1-\beta\cos\theta_{tf})^2}
\int d\xi\, \xi\, F(\xi)
=\frac{(2\pi)^2 (1-\beta^2)}{(1-\beta\cos\theta_{tf})^2}
{\mit\Gamma},
\]
eventually we obtain
\begin{equation}
\frac{d\sigma}{d{\mit\Omega}_f}
=\frac1{2\pi}B_f \int d\tilde{\mib{p}}_t
\frac{d\sigma}{d\tilde{\mib{p}}_t}(s_t=n)
\frac{1-\beta^2}{(1-\beta\cos\theta_{tf})^2}.
\label{A-dis1}
\end{equation}
Note that there are only two possible ways that the structure
of the top-quark decay vertex could influence the distribution:
\\ \ \
i) through the width $d{\mit\Gamma}/d\tilde{\mib{p}}_f$,\ \
ii) through the effective polarization vector $n$.\\
Therefore we conclude that {\it if the polarization vector $n$
(i.e. $\alpha^f$) depends neither on $E_f$ nor on anomalous
top-quark-decay vertices, the angular distribution
$d\sigma/d{\mit\Omega}_{f}$ is not altered by those anomalous
vertices except for possible trivial modification of the branching
ratio $B_f$}. Furthermore, if we focus on the single standard-decay
channel $t\to bW^+\to b {\ell}^+ \nu_{\ell}$, even that dependence
disappears. This result is never a trivial matter since the
lepton-energy distribution, e.g., does depend on the anomalous
decay parameters even when $n$ satisfies the above
conditions \cite{Grzadkowski:1996kn}.

Let us stress here the difference between this conclusion and
those in \cite{Grzadkowski:2000iq,Grzadkowski:2001tq}: If we
focus only on the decay mode $t\to bW^+\to b {\ell}^+ \nu_{\ell}$,
the calculations in \cite{Grzadkowski:2000iq,Grzadkowski:2001tq}
can of course give the same result. However we cannot say thereby
anything on other decay modes since we fully used the explicit
decay-width formula for that mode there to get the results. On
the other hand, the present formalism can do. Indeed, there
could be infinite other decay patterns (although 
$t\to bW^+\to b {\ell}^+ \nu_{\ell}$ will be the leading channel).
That is, if we extend the Higgs sector, the top-quark could
decay via charged Higgs exchange. If we introduce effective
four-fermion interactions, then it would decay through a contact 
$[tb{\ell} \nu]$ coupling. Even within the standard model, $t\to
bW\to b{\ell} \nu +\ arbitrary\ number\ of\ photons$ could occur.
The present formula is applicable to any of them. This comparison
clearly shows that our present formalism is the full important
generalization of our previous works.

Finally, we consider the structure of $n$ for the main-decay mode
$t \to b {\ell}^+ \nu_{\ell}$ focusing on $f={\ell}^+$ and $b$.
Within the SM, it was found in Ref.\cite{Arens:1992wh} that
\begin{equation}
\alpha^{{\ell}^+}=1\ \ \ {\rm and}\ \ \
\alpha^{b}=(2M_W^2 -m_t^2)/(2M_W^2 +m_t^2).
\end{equation}
Using the most general covariant $tbW$ coupling
\begin{equation}
{\mit\Gamma}^{\mu} \propto
\bar{u}(p_b)\biggl[\,\gamma^{\mu}(f_1^L P_L +f_1^R P_R)
-{{i\sigma^{\mu\nu}k_{\nu}}\over M_W}
(f_2^L P_L +f_2^R P_R)\,\biggr]u(p_t),  \label{tbW}
\end{equation}
where $P_{L/R}=(1\mp\gamma_5)/2$ and $k$ is the momentum of $W$, we
have confirmed in \cite{Grzadkowski:2000iq} that $\alpha^{{\ell}^+}$
remains unchanged while $\alpha^b$ receives corrections proportional
to ${\rm Re}(f_2^R)$. In those calculations, all the fermions except
$t$ were treated as massless, the narrow-width approximation was
adopted also for the decaying $W$, and only the
$[$SM$]$-$[$non-SM$]$ interference terms were taken into account.
Thus, $n$ was indeed  $E_f$-independent for those cases within our
approximations, and the leptonic angular distribution cannot depend
on the decay interactions even for the general $tbW$ vertex. 
Since the depolarization factor for $f=b$ was found to be sensitive
to non-SM interactions already for $m_b=0$,
hereafter we shall focus only on the leptonic distribution.

In the above calculations, we have used the following spinor $B$, 
defined below eq.(\ref{n-vec}), for the interaction specified in
(\ref{tbW}):
\begin{eqnarray}
&&B\propto \biggl[\,\gamma^{\mu}(f_1^{L*} P_L +f_1^{R*} P_R)
+{{i\sigma^{\mu\nu}k_{\nu}}\over M_W}
(f_2^{L*} P_R +f_2^{R*} P_L)\,\biggr]u(p_b) \non\\
&&\phantom{B=}
\times
\bar{v}(p_{\ell})\gamma_{\mu}P_L u(p_{\nu})
\,\delta(k^2 -M_W^2)\, ,  \label{B-spinor}
\end{eqnarray}
while the phase-space
element $d{\mit\Phi}$ needed in (\ref{n-vec}) is
\begin{equation}
d{\mit\Phi}=d\tilde{\mib{p}}_b d\tilde{\mib{p}}_{\nu}
\delta^4(p_t -p_b -p_{\ell} -p_{\nu}) .
\label{dPhi}
\end{equation}
Let us now test the relevance of the approximations so far adopted.
First the $b$-quark mass: It is seen from the form of $B$ that if
$m_b=0$ then only $f_2^R$ term can interfere with the leading $V-A$
from factor. However, for $m_b\neq 0$ all the form factors do
contribute. Nevertheless, we have found via explicit analytical
calculations that {\it the depolarization factor for $f={\ell}^+$
still remains unchanged:
$\alpha^{{\ell}^+}=1$ even when $m_b$ is not neglected.} So, when
only interference terms are kept, $\alpha^{{\ell}^+}$ depends neither
on $E_{{\ell}^+}$ nor on the anomalous couplings regardless what
bottom-quark mass was employed! On the other hand, 
however, we have noticed that $\alpha^{{\ell}^+}$ does receive non-SM
corrections of the order of $[{\rm \mbox{non-SM}}]^2$ not only when
eq.(\ref{tbW}) is used for the $tbW$ vertex but also when
similar anomalous terms were introduced into the ${\ell}\nu_{\ell}W$
vertex. The narrow-width approximation for the decaying $W$ was also
found to be inevitable for $\alpha^{{\ell}^+}=1$. That is, if we
replace $\delta(k^2 -M_W^2)$ in eq.(\ref{B-spinor}) by the squared
$W$ propagator with a finite width, $\alpha^{{\ell}^+}$ does not stay
unchanged even if only interference terms are kept.

There are a few remarks here in order:
\bit
\item The angular dependence seen in eq.(\ref{A-dis1}) cannot have
any dynamical origin as we have never specified top-quark
interactions. The angular distribution
$d{\mit\Gamma}/d\tilde{\mib{p}}_f$ entering eq.(\ref{KST1}) is
isotropic in the top-quark rest frame as top quark is unpolarized.
Therefore the dependence on $\theta_{tf}$ of the integrand
(\ref{A-dis1}) is just a remnant of the Lorentz
transformation from the top-quark rest frame to the LAB frame:
\beq
\frac{d{\mit\Gamma}}{d\cos\theta_{tf}}
=\frac{1-\beta^2}{(1-\beta\cos\theta_{tf})^2}
\frac{d{\mit\Gamma^\star}}{d\cos\theta^\star},
\eeq
where $d \mit\Gamma^\star /d\cos\theta^\star$ is the constant
distribution defined in the top-quark rest frame.
\item Our derivation of the angular distribution (\ref{A-dis1})
applies to any top-quark production process, including pair
and single productions at both $\epem$ and hadronic machines (or
$\gamma\gamma$ collisions enabled by laser-electron/positron backward
scatterings) in the open-top region. For $\epem$ collisions the absolute
value of top-quark
momentum is fixed by $\beta^2=1-4m_t^2/s$ and eq.(\ref{A-dis1})
reduces to
\beq
\frac{d\sigma}{d{\mit\Omega}_f} = \frac{2m_t^2}{\pi s}B_f
\int d{\mit\Omega}_t \frac{d\sigma}{d{\mit\Omega}_t}(s_t=n)
\frac1{(1-\beta\cos\theta_{tf})^2},
\eeq
which is what we have derived in \cite{Grzadkowski:2001tq}
including beyond-the-SM interactions.\footnote{The analogous SM
result has been found in Ref.\cite{Arens:1992wh,Arens:1992fg}.}
On the other hand, the distribution in the CM frame of hadron-hadron
collisions has some additional factors since the hadron CM frame and
the parton CM frame are different from each other and they are
connected through Lorentz transformation. However, any Lorentz boost
can never produce anomalous-decay-parameter dependence. So, if
$d\sigma/d\cos\theta$ in the parton-CM frame is free from the non-SM
form factors, then the one in the hadron-CM frame is also free from
them. Consequently, our decoupling theorem holds in hadron-hadron
collisions, too.
\eit

In conclusion, we have investigated  the angular distribution of a
secondary particle $f$ (without concentrating only on
$f = {\ell}^+, b$) in processes like $1+2 \to t + \cdots$ followed
by $t \to f + \cdots$ neglecting the $f$ mass and applying the
narrow-width  approximation for the decaying top quark. It has been shown
that if the effective polarization vector $n$ contains neither non-SM
top-quark couplings nor $E_f$ (as is the case for $f={\ell}^+$ within our
approximation) then the whole angular distribution of $f$ has no
non-SM top-quark-decay contributions. Non-standard corrections can
enter the leptonic distribution only through modification of $n$ as
corrections to the narrow-width approximation, $m_{\ell} \neq 0$
terms or contributions quadratic in non-SM vertices. However, if the
polarization vector $n$ does contain non-SM contributions (as is the
case for $f=b$), then the angular distribution of $f$ receives extra
correction from the top-quark-decay vertices only through the
production cross section of the polarized top quark.

We emphasize that our conclusions concerning the decoupling holds for
any possible top-quark production mechanism even if the bottom-quark
mass is not neglected. Therefore unknown top-quark couplings that
parameterize the angular distribution of $f$ in the case of SM
polarization vector (e.g. $f={\ell}^+$) are reduced to those that
influence the production process. Since fewer unknown parameters lead
to higher precision for their determination, we believe that the
angular distribution will be useful while testing top-quark couplings
at future colliders.

\vspace*{0.6cm}
\centerline{ACKNOWLEDGMENTS}

\vspace*{0.3cm}

B.G. thanks Jos\'e Wudka for interesting discussions. B.G. is also
indebted to U.C. Riverside for its warm hospitality extended to him
while this work was being performed. Z.H. is grateful to Y. Sumino
for valuable conversations on the QCD corrections in the threshold
region. This work is supported in part
by the State Committee for Scientific Research under grant
5~P03B~121~20 (Poland) and the Grant-in-Aid for Scientific Research
No.13135219 from the Japan Society for the Promotion of Science.

\vspace*{0.6cm}

\end{document}